\documentclass[12pt,preprint]{aastex}

\def\ifundefined#1{\expandafter\ifx\csname#1\endcsname\relax}

\def\la{\mathrel{\hbox{\rlap{\hbox{\lower4pt\hbox{$\sim$}}}\hbox{$<$}}}}
\def\ga{\mathrel{\hbox{\rlap{\hbox{\lower4pt\hbox{$\sim$}}}\hbox{$>$}}}}

\newcommand{\be}{\begin{eqnarray}}
\newcommand{\ee}{\end{eqnarray}}

\ifundefined{ensuremath}\def\ensuremath#1{\relax\ifmmode{#1}}
\else${#1}$\fi\else\relax\fi
\ifundefined{nuc}\def\nuc#1#2{\relax\ifmmode{}^{#1}{\protect\textrm{#2}}
\else${}^{#1}$#2\fi}\else\relax\fi

\newcommand{\etal}{et al.}

\newcommand{\kmps}{\ensuremath{\mathrm{km}~\mathrm{s}^{-1}}}

\newcommand{\msol}{\ensuremath{{\mathrm{M}_\odot}}}

\def\ang{\ensuremath{\mathrm{\AA}}}
\def\Tmod{\ensuremath{T_{\mathrm{model}}}}

\def\teff{\ensuremath{T_{\mathrm{model}}}}
\def\tstd{\ensuremath{\tau_{\mathrm{std}}}}

\newcommand{\vno}{\ensuremath{v_0}}

\newcommand{\phx}{\texttt{PHOENIX}}

\newcommand{\halpha}{H$\alpha$}
\newcommand{\hbeta}{H$\beta$}

\shortauthors{Baron, E. et~al.}
\shorttitle{Type Ia Supernova Spectral Line Ratios}

\begin{document}

\title{Reddening, Abundances, and Line Formation in SNe~II}

\author{ 
  E.~Baron,\altaffilmark{1,2}\email{baron@nhn.ou.edu}
  David
Branch,\altaffilmark{1}\email{branch@nhn.ou.edu}  and  Peter
H.~Hauschildt\altaffilmark{3}\email{yeti@hs.uni-hamburg.de}
 }

 \altaffiltext{1}{Homer L.~Dodge Department of Physics and Astronomy,
   University of Oklahoma, 440 West Brooks, Rm.~100, Norman, OK
   73019-2061, USA}

\altaffiltext{2}{Computational Research Division, Lawrence Berkeley
  National Laboratory, MS 50F-1650, 1 Cyclotron Rd, Berkeley, CA
  94720 USA}

\altaffiltext{3}{Hamburger Sternwarte, Gojenbergsweg 112,
21029 Hamburg, Germany}

\begin{abstract}
We present detailed NLTE spectral synthesis models of the Type II supernova
2005cs, which occurred in M51 and for which the explosion time is well
determined. We show that previous estimates for the reddening were
significantly too high and briefly discuss how this will effect the inferred
progenitor mass. We also show that standard CNO-burning enhanced
abundances require far too large an oxygen depletion, although there is
evidence for a single optical N~II line and the sodium
abundance shows clear evidence for enhancement over solar both as expected
from CNO processing. 
Finally we calculate a distance
using the SEAM method. Given the broad range of distances to M51 in
the literature, the determination of a distance using Cepheid
variables would be quite valuable.
\end{abstract}


\section{Introduction} 

Supernova 1987A (SN~1987A) confirmed our basic theoretical
understanding of a Type II supernova (SN~II) as the core collapse of a
massive star, which leaves behind a compact object (neutron star or
black hole) and expels the outer mantle and envelope into the
interstellar medium \citep[see][and references
therein]{sn87arev89}. While the explosion mechanism itself remains a
subject of active research, the propagation of the shock wave through
the mantle and envelope is reasonably well understood. Theoretical
models for the light curve do a good job of reproducing the
observations \citep{blinn87a99a,blinn87a00} and detailed radiation
transport calculations verify that these models reproduce the observed
spectra from the UV to the IR
\citep{mitchetal87a01,mitchetal87a02}. Unlike Type Ia supernovae
(SNe~Ia) where the actual compositions as a function of velocity are
an important subject of current research
\citep{fisher91T99,stehle02bo05}, the compositions of the envelopes of
red or blue supergiant stars is primarily hydrogen and helium and thus
we show that the primordial compositions can be determined by the detailed
spectral modeling of observed SN~II spectra. Of particular interest is
the effect of CNO processing on the composition of the ejected
material.

SNe~II have a very large spread in their intrinsic brightness, from
the very dim SN~1987A, to the exceedingly bright SNe~1979C and
1997cy. The observed spread in intrinsic luminosity is greater than a
factor of 500. This is not surprising given the fact that the
progenitors span a wide range of initial stellar masses, possible
binary membership, and prior star formation histories. Clearly, SNe~II
do not meet the astronomers requirement of being a \emph{standard
  candle}; however, since our models determine the stellar
compositions and total reddening this provides evidence that their
atmospheres can 
be well understood.  These results increase the attractiveness of
SNe~II as cosmological probes and in the future SNe~II will become
complementary with SNe~Ia as distance indicators. Both the SEAM
(Spectral-fitting Expanding Atmosphere Method)
\citep[see][and references therein]{b93j4,b94i1,mitchetal87a02} and
the EPM (Expanding Photosphere Method)
\citep[see][]{baadeepm,kkepm,branepm,eastkir89,skeetal94,esk96,hamuyepm01,leonard99em02} 
methods for determining distances to SNe~II (and other supernova types
for SEAM) depend on the ability to model the spectral energy
distribution (SED) of the supernova atmosphere accurately. Clearly,
quantities like the composition and total reddening play a role in the
output SED and thus the work presented here helps to place both these
methods on firmer footing.

While the average SN~II has a luminosity several times less than a
typical SN~Ia, current 
ground-based searches and proposed future space-based searches for
supernovae will easily detect these objects at cosmologically
interesting distances. Using an empirical relation between brightness
and expansion velocity on the plateau portion of the light curve
\citet{nugent06}  were able to construct a Hubble diagram using 5
SNe~II out to a redshift $z=0.3$.

The explosion mechanism is thought to result in a non-spherically
symmetric shock-wave \citep[see for example][]{lbw70,khok_jet99} and this has
been confirmed by the detection of significant polarization in the
spectra of SNe~II \citep{wang_pol96,leonmd00,leonard04dj06}. Nevertheless, the
large expansion of the ejecta before the supernova becomes visible
leads to significant ``sphericalization'' \citep{chev84}, and
asphericity effects may only be important at late times, or in SNe~II
that are strong circumstellar interacters (SNe~IIn). These supernovae
are observationally clearly distinguishable  from ``normal'' Type II
supernovae.

Calculations \citep{miralda97,DF99,livyoung00} indicate that SNe~II may in
fact be the first stellar objects visible in the universe and hence
they serve as important probes for the star formation rate, the rate
of chemical evolution by measuring their primordial abundances, and
directly for the cosmological parameters.  We
show that the primordial abundances of the progenitor star can be
determined by detailed synthetic spectral modeling. In this work we do
not seek to determine exact nucleosynthetic yields, but rather show
that this is feasible given a large grid of models that will become
available as more supernovae are analyzed and as computational
resources become more plentiful.

\section{Calculations}

We have chosen to model SN~2005cs in detail because there are several
high signal to noise spectra and $UBVRI$ photometry very near the time
of explosion as well 
as at later times. SN~2005cs was discovered in M51 (NGC~5194) on 2005
June 28.905 \citep{kloehriau05cs05}, the earliest detection was that
of M.~Fiedler on 2005 June 27.91 \citep{pasto05cs06} and nothing was
visible on June 20.6 \citep{kloehriau05cs05}. Moreover no other
detections were found on June 26 by other amateur observers
\citep{pasto05cs06}, thus the explosion date is known to within about
1 day. Using the rather long distance of \citet[][see
below]{feldetal97}, \citet{pasto05cs06} classified SN~2005cs as
moderately under-luminous compared to typical SNe~IIP and in the class of
SN~1997D and SN~1999br. In addition a progenitor has been identified
for SN~2005cs \citep{maund05cs05,li05cs06} with a mass of about 10~\msol.

The calculations were performed using the multi-purpose stellar
atmospheres program \phx~{version \tt 14}
\citep{hbjcam99,bhpar298,hbapara97,phhnovetal97,phhnovfe296}.
\phx\ solves the radiative transfer equation along characteristic rays
in spherical symmetry including all special relativistic effects.  The
non-LTE (NLTE) rate equations for many ionization states are solved
including the effects of ionization due to non-thermal electrons from
the $\gamma$-rays produced by the  radiative decay of $^{56}$Ni, which
is produced in the
supernova explosion.  The atoms and ions calculated in NLTE are: H~I,
He~I--II, C~I-III, N~I-III, O~I-III, Ne~I, Na~I-II, Mg~I-III, Si~I--III,
S~I--III, Ca~II, Ti~II, Fe~I--III, Ni~I-III, and Co~II. These are all the
elements whose features make important contributions to the observed
spectral features in SNe~II.

Each model atom includes primary NLTE transitions, which are used to
calculate the level populations and opacity, and weaker secondary LTE
transitions which are included in the opacity and implicitly
affect the rate equations via their effect on the solution to the
transport equation \citep{hbjcam99}.  In addition to the NLTE
transitions, all other LTE line opacities for atomic species not
treated in NLTE are treated with the equivalent two-level atom source
function, using a thermalization parameter, $\alpha =0.05$.  The
atmospheres are iterated to energy balance in the co-moving frame;
while we neglect the explicit effects of time dependence in the
radiation transport equation, we do implicitly include these effects,
via explicitly including the rate of gamma-ray deposition in the generalized
equation of radiative equilibrium and in the rate equations for the
NLTE populations.

The models are parameterized by the time since explosion and the
velocity (\vno) where the continuum optical depth in extinction at
5000~\ang\ ($\tstd$) is unity, which along with the density profile
determines the radii. This follows since the explosion becomes
homologous ($v \propto r$) quickly after the shock wave traverses the
entire star. The density profile is taken to be a power-law in radius:
\[ \rho \propto r^{-n} \]
where $n$ typically is in the range $6-12$. Since we are only modeling
the outer atmosphere of the 
supernova, this simple parameterization agrees well with detailed
simulations of the light curve \citep{blinn87a00} for the relatively
small regions of the ejecta that our models probe.

Further fitting parameters are the model temperature $\Tmod$, which is
a convenient way of parameterizing the total luminosity in the
observer's frame. We treat the $\gamma$-ray deposition in a simple
parameterized way, which allows us to include the effects of nickel
mixing which is seen in nearly all SNe~II.
Detailed fitting of a time series of the observed spectra determines
all the parameters.

\section{Reddening}

\begin{figure}
\leavevmode
\begin{center}
\includegraphics[width=0.75\textwidth,angle=90]{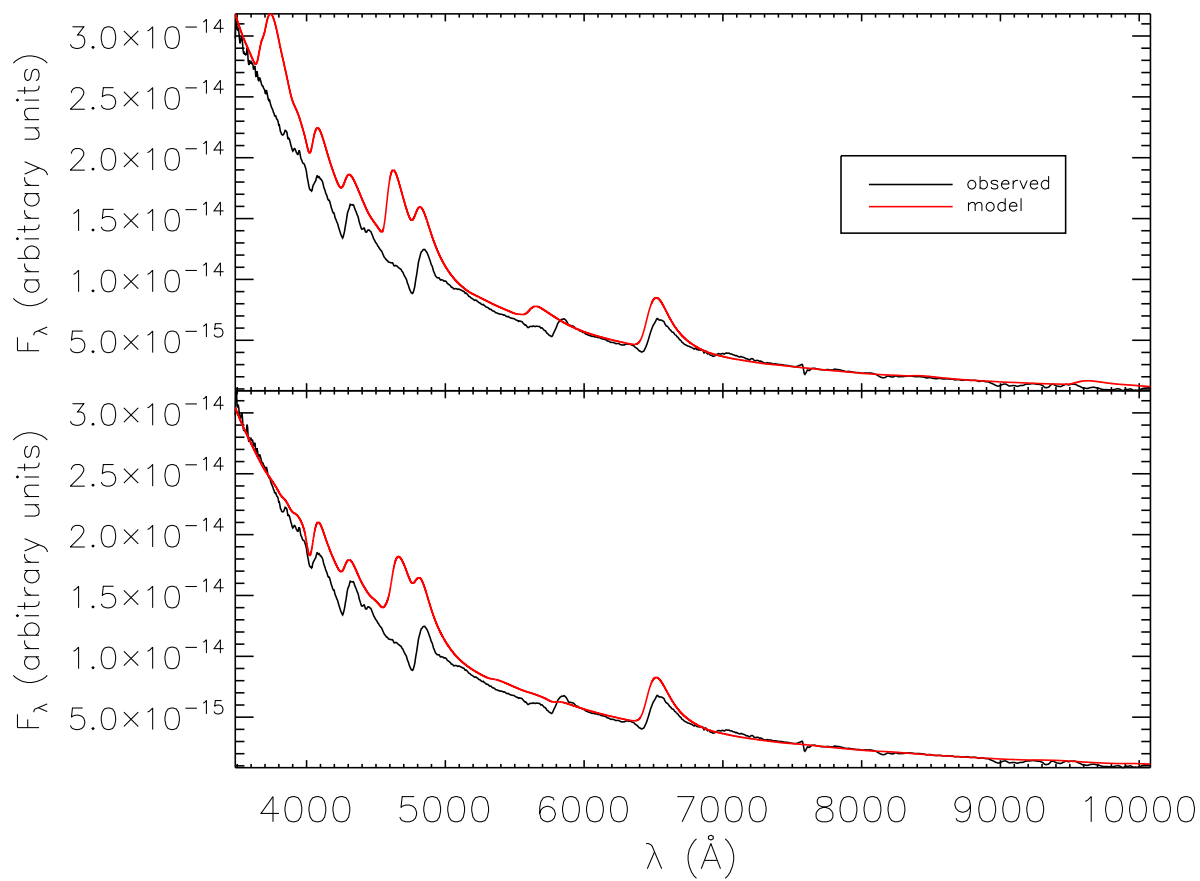}
\end{center}
\caption{\label{fig:t18_ebmv11}. The observed spectrum on Jul 02,
  2005 is compared to a model spectrum using solar abundances (top
  panel) and CNO enhanced abundances (bottom panel). The
  observed spectrum has been 
  dereddened using $E(B-V) = 0.11$. In this and subsequent plots the
  model and observations have been normalized at 8000~\ang.}
\end{figure}

\begin{figure}
\leavevmode
\begin{center}
\includegraphics[width=0.75\textwidth,angle=90]{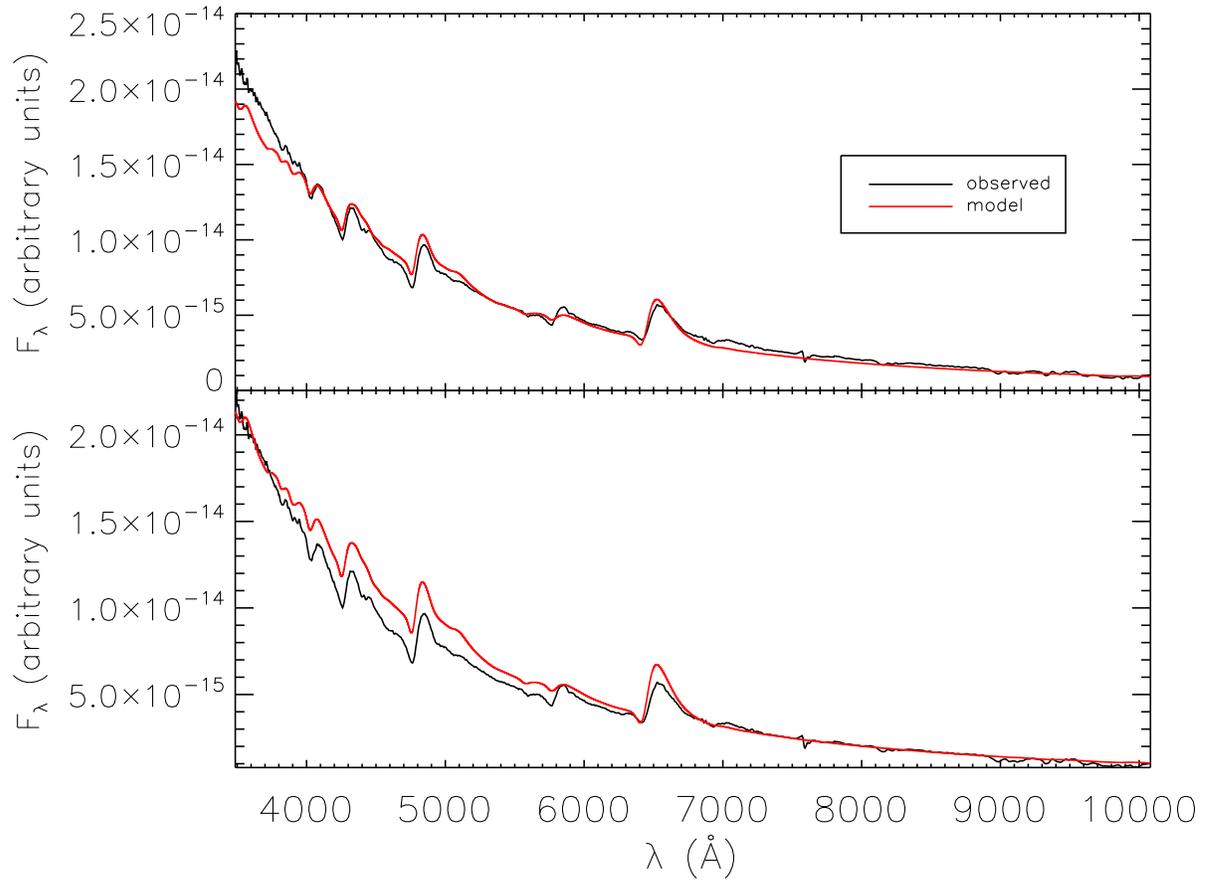}
\end{center}
\caption{\label{fig:t12_ebmv035}. The observed spectrum on Jul 02,
  2005 is compared to a model spectrum using solar abundances (top
  panel) and CNO enhanced abundances (bottom panel). The
  observed spectrum has been 
  dereddened using $E(B-V) = 0.035$.}
\end{figure}

Determining the extinction to SNe~II is difficult, since they are such
a heterogeneous class, it is difficult to find an intrinsic feature
in the spectrum or light curve that can be used to find the parent
galaxy extinction. \citet{sn93w103,bsn99em00} found that the Ca II H+K
lines can be used as a temperature indicator in modeling very early
observed spectra. This suggests that detailed modeling of SNe~IIP
spectra can be used to determine the total reddening to the supernova. For SN~2005cs the reddening has been estimated in a
variety of ways. \citet{maund05cs05} used the relationship between the
equivalent width of the Na~I~D interstellar absorption line to obtain a
color excess of $E(B-V) = 0.16$, as well as the color magnitude
diagram of red supergiants within 2 arcsec of SN~2005cs to obtain
$E(B-V) = 0.12$, and their final adopted $E(B-V) =
0.14$. \citet{li05cs06} noted the large scatter in the relationships
for equivalent width of the Na~I~D line, obtaining a range of $E(B-V)
= 0.05-0.20$. Assuming that the color evolution of SN~2005cs is
similar to that of SN~1999em, they found $E(B-V) = 0.12$. Also using
the Na~I~D  line \citet{pasto05cs06} found $E(B-V) = 0.06$, but noting
the uncertainty and comparing with the work of other authors they
adopted $E(B-V) = 0.11$. We began our work by adopting the reddening
estimate of \citet{pasto05cs06}, since our spectra were obtained from
these authors. 
Figure~\ref{fig:t18_ebmv11} (top panel) shows our best fit using solar
abundances \citep{GS98} where the observed spectrum has been
dereddened using the 
reddening law of \citet{card89} and
$R_V=3.1$. Figure~\ref{fig:t18_ebmv11} (bottom panel) shows the same for the case
of CNO enhanced abundances \citep{dessart05a,prantzos86}. From the 
figure it is evident that the 
region around H$\beta$\ is very poorly fit, there is a strong feature
just to the blue of \hbeta\ and \hbeta\ itself is far too weak. We
attempted to alter the model in a number of ways, changing the density
profile, velocity at the photosphere, and gamma-ray deposition in
order to strengthen \hbeta, however we were unable to find any set of
parameters that would provide a good fit to \hbeta\ (and the rest of
the observed spectrum) with this choice of
reddening. These models have \teff=18000~K, \vno = 6000~\kmps, and $n =
8$. After careful analysis the strong feature too the blue of \hbeta\
is due to C~III and N~III lines that become strong at these high
temperatures. Figure~\ref{fig:t12_ebmv035} 
shows that if \teff\ is reduced to 12000~K and the color excess is
reduced to the galactic foreground value of $E(B-V) = 0.035$
\citep{schlegelred98} the fit is significantly improved. Our value of
$E(B-V) = 0.035$ is in agreement within the errors of the lower values
found by \citet{li05cs06} and \citet{pasto05cs06}. Clearly, the Na D
interstellar absorption indicates that there was some dust in the host
galaxy, but we have chosen the value that gave the best fit for this
epoch which was no dust extinction in the host galaxy. Given our
modeling uncertainties, $E(B-V)=
0.035-0.05$, but we will adopt the value of 0.035 for the rest of this
work. Note that this lower value of the extinction will somewhat lower
the inferred mass of the progenitor found by \citet{maund05cs05}
and \citet{li05cs06}, but other uncertainties such as distance and
progenitor metallicity also play important roles in the uncertainty of
the progenitor mass. Comparing with the published results, the
difference due to our distance and reddening estimates on the value of
the progenitor 
mass obtained is quite small. Using the value of
$R_V - R_I = 1.28$ and $R_I = 1.89$ \citep{TSR03} the change in $I$ is
$\sim 0.15$~mag and $(V-I) \sim 0.13$~mag. Our derived distance is
consistant with that adopted by \citet{li05cs06} and only somewhat
smaller than that of \citet{maund05cs05}. 

Clearly the bluest part of the continuum is
better fit with the \teff=18000~K models than with the \teff=12000~K
models. We did not attempt to fine tune our results to perfectly fit
the bluest part of the continuum since the flux calibration at the
spectral edges is difficult and it represents our uncertainty in
\teff\ 
and $E(B-V)$. None of the above models have He~I $\lambda 5876$ strong
enough. It is well known that Rayleigh-Taylor instabilities lead to
mixing between the hydrogen and helium shells thus our helium
abundance is almost certainly too low, but we will not explore helium
mixing further in this work. 
\clearpage

\clearpage

\section{Abundances}

Typically, the approach to obtaining abundances in differentially
expanding flows has been through line identifications. This method has
been successful using the SYNOW code 
\citep[see][and references therein]{branchcomp105,branchcomp206} as
well as the work of Mazzali and collaborators \citep[for
example][]{stehle02bo05}.  
Nevertheless, line identifications do not provide direct information
on abundances, which are what are input into stellar evolution
calculations and output from hydrodynamical calculations of supernova
explosions and nucleosynthesis. Line identifications are subject to
error in that there may be another candidate line that is not
considered in the analysis or there may be two equally valid possible
identifications, the classic being He~I $\lambda 5876$ and Na~I D,
which have very similar rest wavelengths and are both expected in
supernovae (and have both been identified in supernovae).

Here we focus on nitrogen and sodium. Massive stars which are the
progenitors of SNe~II are expected to undergo CNO processing at the
base of the hydrogen envelope followed by mixing due to dredge up and
meridional circulation. This would lead to enhanced nitrogen and
sodium and depleted carbon and oxygen. For our CNO processed models we
take the abundances used by \citet{dessart05a}. However, significant
mixing of the hydrogen and helium envelope is expected to occur during
the explosion due to Rayleigh-Taylor instabilities and mass loss will
occur during the pre-supernova evolution. Studies of the evolution of
massive stars show that for an 11~\msol\ model the mass fractions of
CNO are very close to solar, with significantly less depletion of
carbon and oxygen than is obtained from CNO processing (A.~Chieffi \&
M.~Tavani, 2006, in preparation). A detailed study of SN~2005cs
spectra using the results of these hydro models will be a sensitive
test of the predictions of stellar evolution theory.

\subsection{Line IDs\label{sec:line_ids}}

\begin{figure}
\leavevmode
\begin{center}
\includegraphics[width=0.75\textwidth,angle=90]{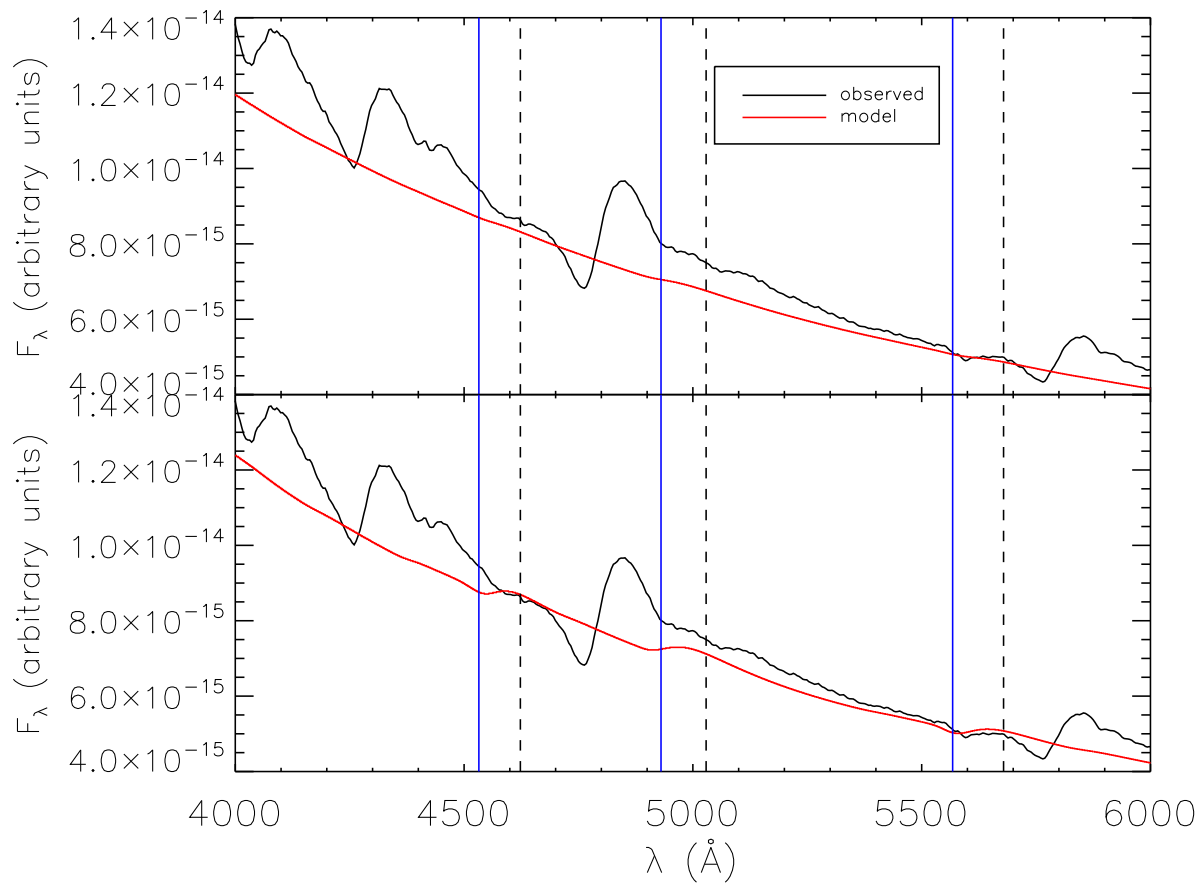}
\end{center}
\caption{\label{fig:nii_single_element} The observed spectrum on Jul
  02, 2005 is compared to a model spectrum where only N~II line
  opacity has been included. The top panel is for solar abundances and
  the bottom panel using CNO enhanced abundances.  The vertical dashed
  lines show the rest wavelength of the N~II optical lines, while the
  solid vertical lines show those same lines blueshifted by
  $\vno=6000$~\kmps.  The observed spectrum has been dereddened using
  $E(B-V) = 0.035$.}
\end{figure}

In detailed line-blanketed models such as the ones presented here line
identifications are difficult since nearly every feature in the model
spectrum is a blend of many individual weak and strong
lines. Nevertheless, it is useful to attempt to understand just what
species are contributing to the variations in the spectra. In order to
do this we produce ``single element spectra'' where we calculate the
synthetic spectrum (holding the temperature and density structure
fixed) but turning off all line opacity except for that of a given
species. Figure~\ref{fig:nii_single_element} shows the single element
spectrum for N~II for our \teff\ = 12000~K models with solar abundances
(top panel)
and CNO enhanced abundances (bottom panel). Clearly the N~II lines are
present in CNO enhanced models and absent in the solar models, but
their effect on the total spectrum is unclear. 

\begin{figure}
\leavevmode
\begin{center}
\includegraphics[width=0.75\textwidth,angle=90]{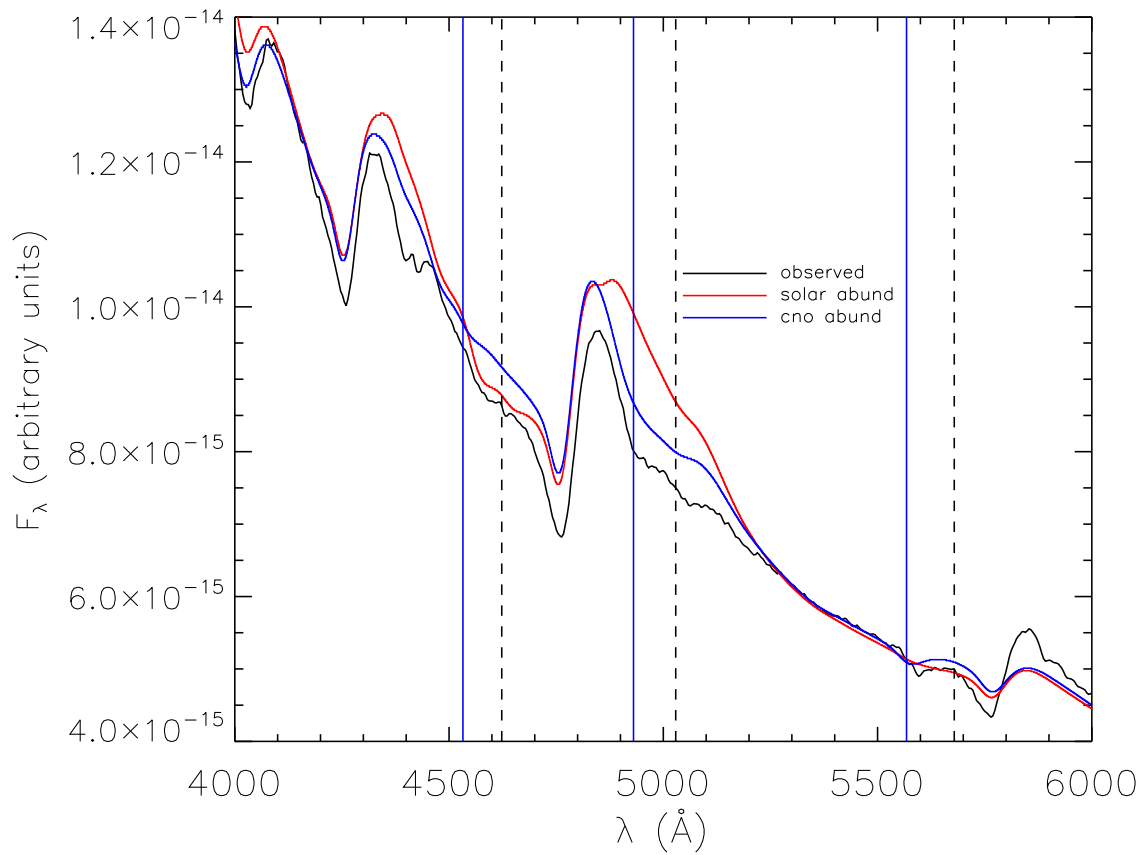}
\end{center}
\caption{\label{fig:cno_v_sol_blowup} The observed spectrum on Jul 02,
  2005 is compared to a model spectrum using CNO enhanced abundances
  and one using solar abundances. The region of N~II lines is shown
  for clarity. The vertical dashed lines show the rest wavelength of
  the N~II optical lines, while the solid vertical lines show those
  same lines blueshifted by $\vno=6000$~\kmps. The
  observed spectrum has been 
  dereddened using $E(B-V) = 0.035$.}
\end{figure}

N~II lines were first identified in SN~II  in SN
1990E \citep{bs90E93}. 
Using SYNOW, \citet{bsn99em00} found evidence for N~II in SN~1999em,
however more detailed modeling with \phx\ indicated that the lines were
in fact due to high velocity Balmer and He~I lines.   Prominent N~II
lines in the optical are N~II $\lambda 4623$, $\lambda 5029$, and
$\lambda 5679$. \citet{dessart05a} found strong evidence for N~II in
SN~1999em. Using SYNOW Elmhamdi \etal\ (in preparation) found
evidence for N~II in SN~2005cs, as did
\citet{pasto05cs06} using the code developed by Mazzali and
collaborators. Figure~\ref{fig:t12_ebmv035}
compares solar abundances to a model with enhanced CNO abundances
\citep{dessart05a,prantzos86}. Figure~\ref{fig:t12_ebmv035} shows that
the CNO enhanced abundances model does somewhat better fitting the
emission peak of \hbeta\ (see below), however the feature to
the blue of \hbeta, 
clearly well-fit in the solar abundance model is completely absent in
the model with enhanced CNO
abundances. Figure~\ref{fig:cno_v_sol_blowup} shows the region where
the optical N~II lines are prominent and in particular, the N~II
$\lambda 5679$ is not quite in the same place as the observed feature
and the two bluer lines have almost no effect. 
\clearpage

\begin{figure}
\leavevmode
\begin{center}
\includegraphics[width=0.75\textwidth,angle=90]{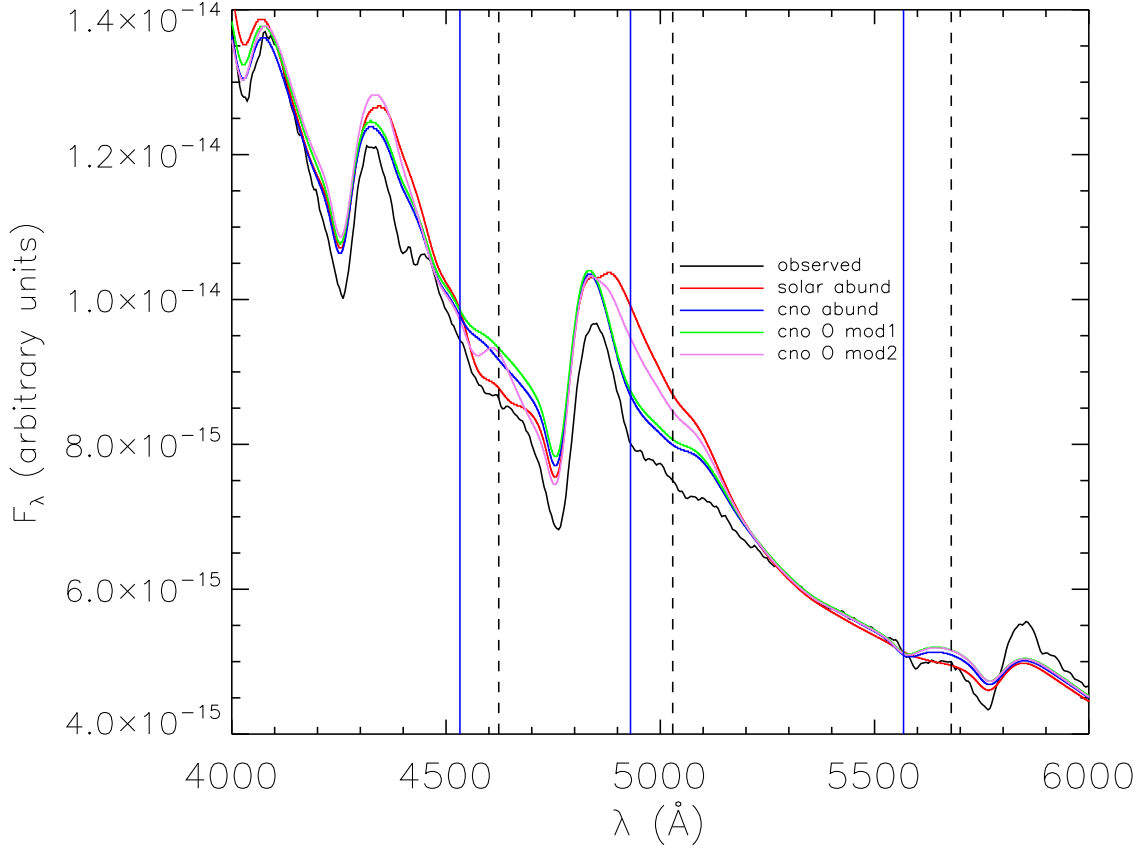}
\end{center}
\caption{\label{fig:cno_modified} The observed spectrum on Jul 02,
  2005 is compared to a model spectrum using CNO enhanced abundances
  and two where the oxygen abundance is enhanced (over the CNO value
  by a factor of 10 and nearly 100) in an attempt to improve the
  \hbeta\ emission feature and still retain the desired O~II feature
  just blueward of \hbeta. The plot shows how complex line
  IDs are. The alteration of the oxygen abundance not only affects the
  features, but the structure of the entire atmosphere itself (since
  oxygen is such a good electron donor). Neither modification is
  satisfactory and clearly more work needs to be done to obtain the
  ``right'' CNO abundances.
  The vertical dashed lines show the rest wavelength of
  the N~II optical lines, while the solid vertical lines show those
  same lines blueshifted by $\vno=6000$~\kmps. The
  observed spectrum has been 
  dereddened using $E(B-V) = 0.035$.}
\end{figure}

\clearpage

We attempted to further improve the fit to the \hbeta\ emission
feature by using the same abundances for C, N, and Na as our standard
CNO processed abundances, but enhancing oxygen by a factor of 10 and
nearly 100 (i.e. using the solar value in the latter
case). Figure~\ref{fig:cno_modified} shows how complex line
  IDs are. The alteration of the oxygen abundance not only affects the
  features, but the structure of the entire atmosphere itself (since
  oxygen is such a good electron donor). Neither modification is
  satisfactory and clearly more work needs to be done to obtain the
  ``right'' CNO abundances. Interestingly, the model with enhanced
  nitrogen and solar oxygen clearly shows the N~II $\lambda 4623$
  line.

\begin{figure}
\leavevmode
\begin{center}
\includegraphics[width=0.75\textwidth,angle=90]{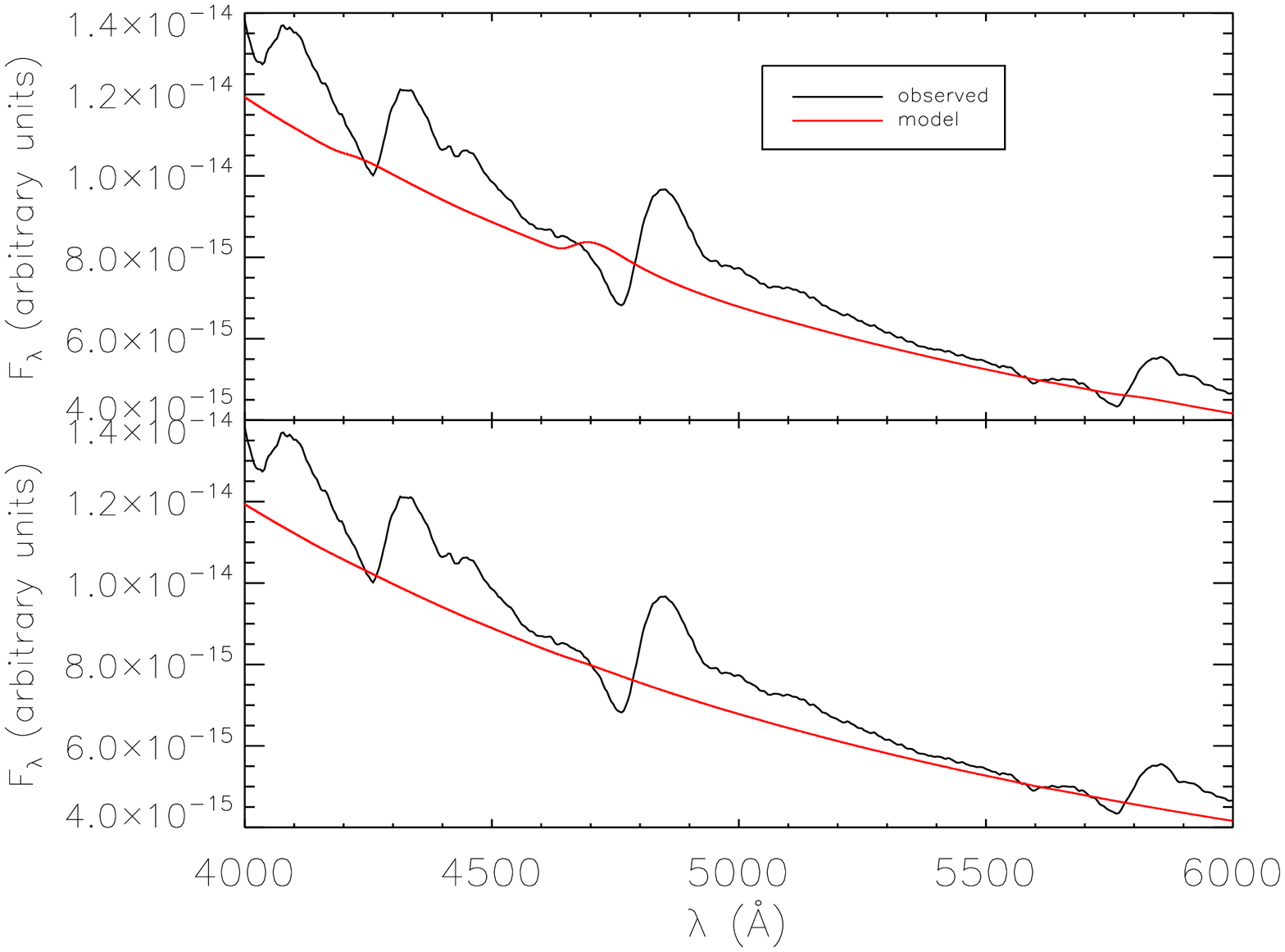}
\end{center}
\caption{\label{fig:cii_single_element} The observed spectrum on Jul 02,
  2005 is compared to a model spectrum where only C~II line opacity
  has been included. The top panel is for solar abundances and the
  bottom panel using CNO enhanced abundances.
  The
  observed spectrum has been 
  dereddened using $E(B-V) = 0.035$.}
\end{figure}

In an attempt to identify the better fit of the feature just to the
blue of \hbeta\ we examined the single element spectrum of C~II
(Fig.~\ref{fig:cii_single_element}) since we considered that the C~II
$\lambda 4619$ line might be playing a role. Again this feature is
clearly present in the solar model and absent in the CNO enhanced
model. 

\begin{figure}
\leavevmode
\begin{center}
\includegraphics[width=0.75\textwidth,angle=90]{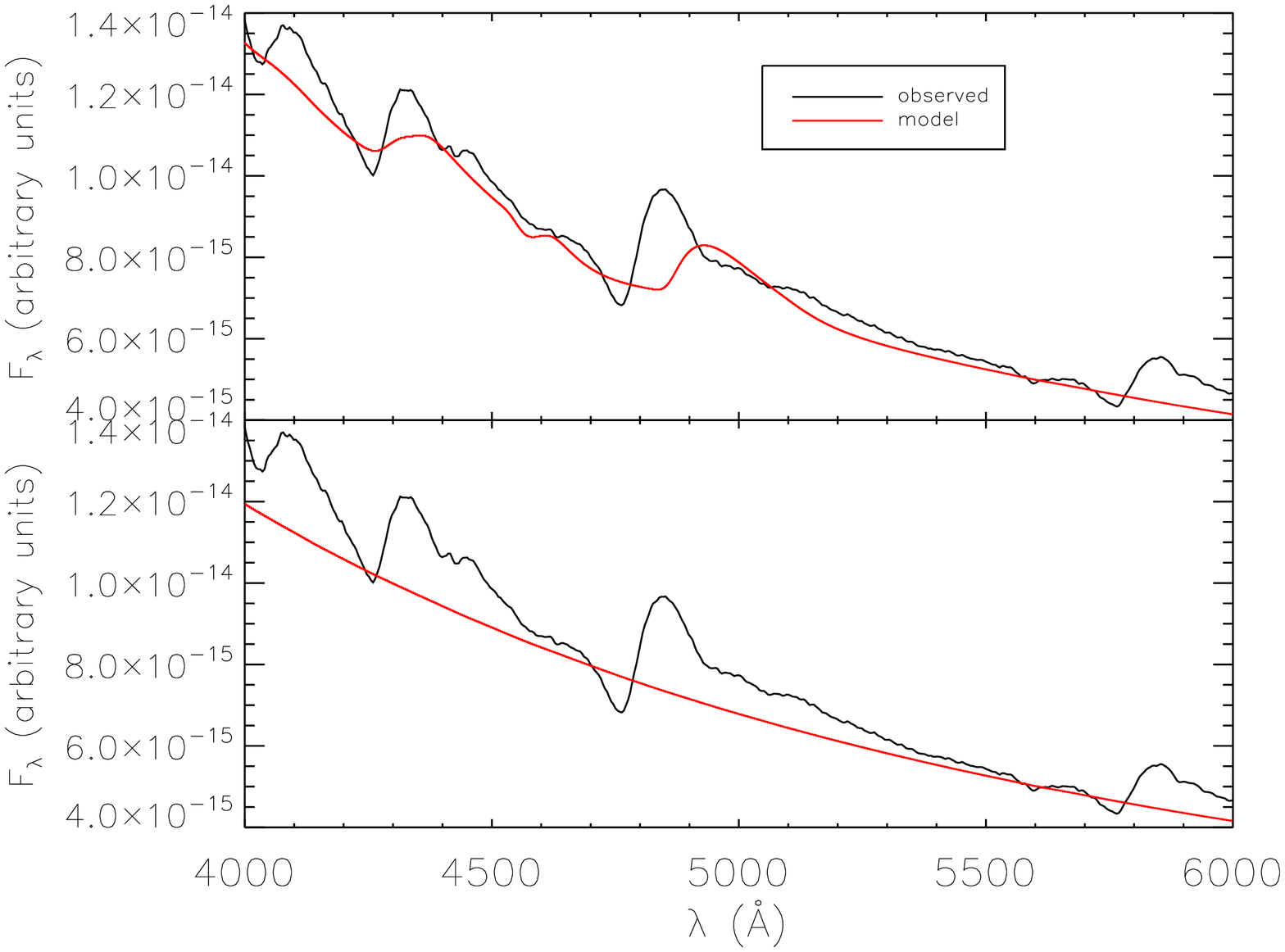}
\end{center}
\caption{\label{fig:oii_single_element} The observed spectrum on Jul 02,
  2005 is compared to a model spectrum where only O~II line opacity
  has been included. The top panel is for solar abundances and the
  bottom panel using CNO enhanced abundances.
  The
  observed spectrum has been 
  dereddened using $E(B-V) = 0.035$.}
\end{figure}

Finally, we examined the lines produced by
O~II. Figure~\ref{fig:oii_single_element} clearly shows that O~II
lines play an important role in producing the observed feature just
blueward of \hbeta. Most likely it is the lines O~II $\lambda 4651.5$
and $\lambda 4698$ which are producing the observed feature. On the
other hand it is also clear that O~II $\lambda 4915$ and $\lambda
4943$ are producing the deleterious feature just to the red of
\hbeta. Since we want somewhat higher oxygen abundances, it was
natural to examine a model using the recently revised solar abundances
of \citet{AGS02}. Compared with our standard solar model abundances
there was not a significant improvement (nor a degradation) in the fit
to favor one over the other.

\begin{figure}
\leavevmode
\begin{center}
\includegraphics[width=0.75\textwidth,angle=90]{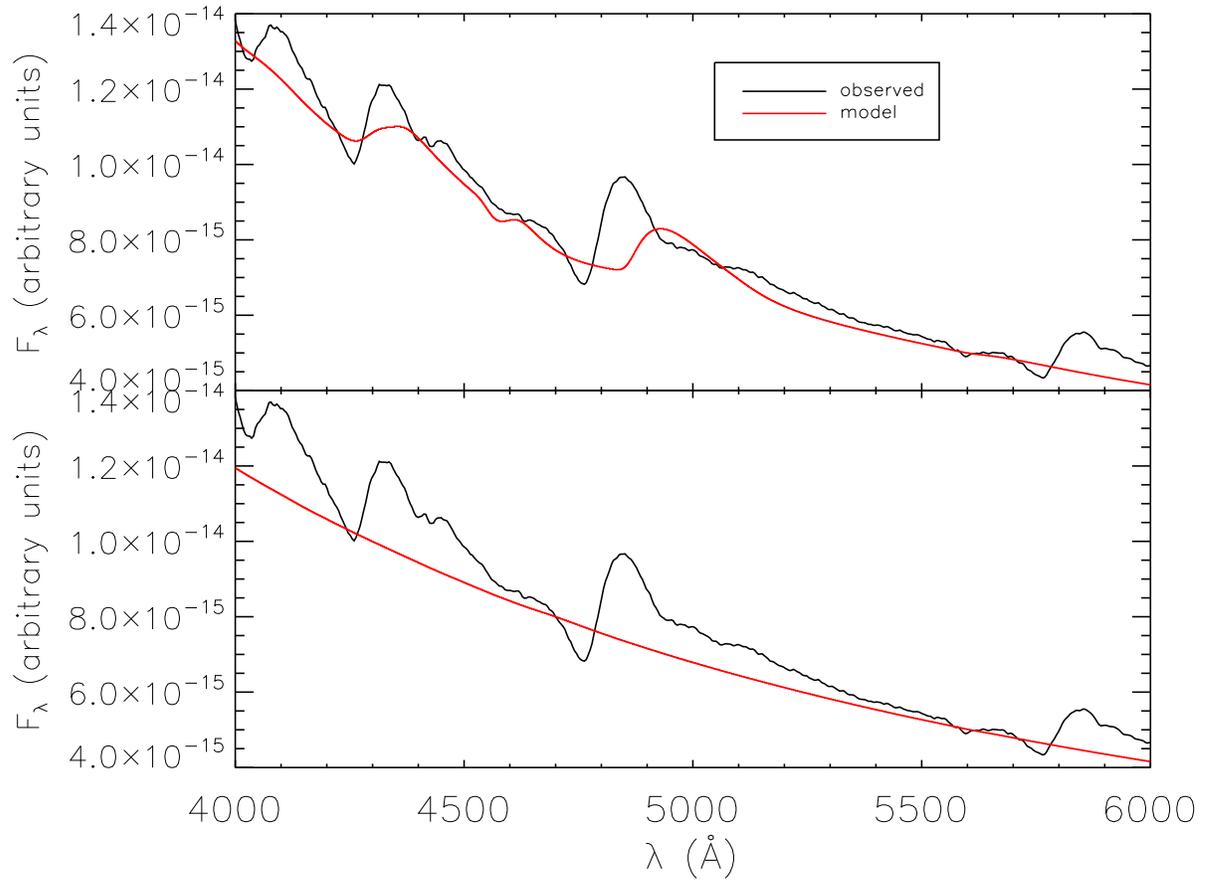}
\end{center}
\caption{\label{fig:cii+oii_single_element} The observed spectrum on Jul 02,
  2005 is compared to a model spectrum where only C~II and O~II line opacity
  has been included. The top panel is for solar abundances and the
  bottom panel using CNO enhanced abundances.
  The
  observed spectrum has been 
  dereddened using $E(B-V) = 0.035$.}
\end{figure}

Figure~\ref{fig:cii+oii_single_element} examines the possibility that
both C~II and O~II contribute to the feature just blueward of
\hbeta. It is obvious that only O~II lines are important for
the structure of this feature. 

Thus, it is clear that the strong depletion of oxygen expected from
CNO processing is not evident, we can not rule out that there is some
enhanced nitrogen in the observed spectra, but the N~II lines don't
seem to form in the right place. However since we are studying simple,
parameterized, homogeneous models this could be an artifact of our
parameterization. Nevertheless, the absorption trough of the feature
that we would like to attribute to N~II $\lambda 5679$ is too fast in
our models, whereas one would expect the CNO processed material with higher
nitrogen abundance
at
the outermost part of the envelope and thus to form at even higher
velocity due to homologous expansion. Clumping could of course change
this simple one-dimensional picture.

\clearpage

\clearpage

\section{Day 17}

\begin{figure}
\leavevmode
\begin{center}
\includegraphics[width=0.75\textwidth,angle=90]{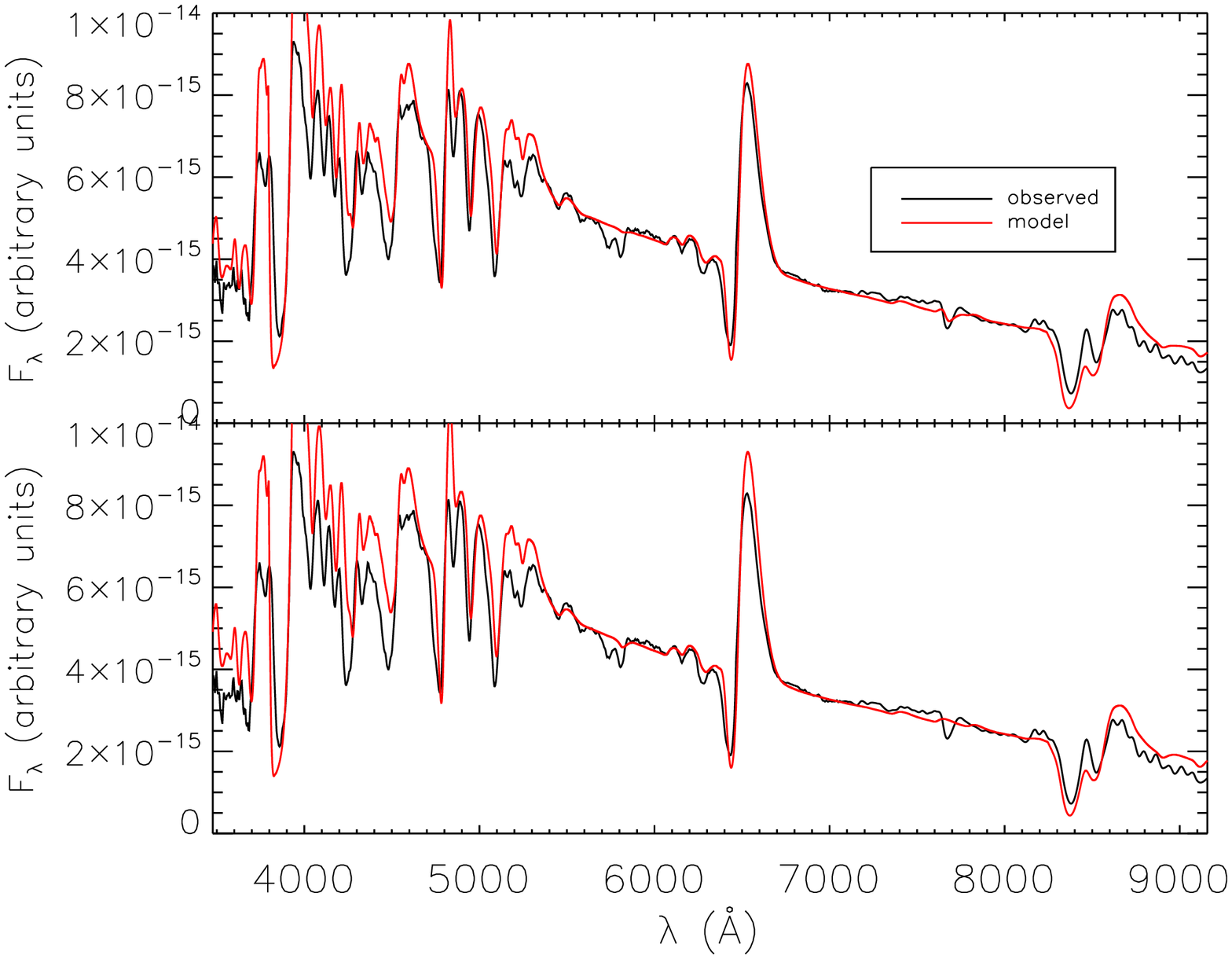}
\end{center}
\caption{\label{fig:d17_std}. The observed spectrum on Jul 14,
  2005 is compared to a model spectrum using solar abundances (top
  panel) and CNO enhanced abundances (bottom panel). The
  observed spectrum has been 
  dereddened using $E(B-V) = 0.035$.}
\end{figure}

\begin{figure}
\leavevmode
\begin{center}
\includegraphics[width=0.75\textwidth,angle=90]{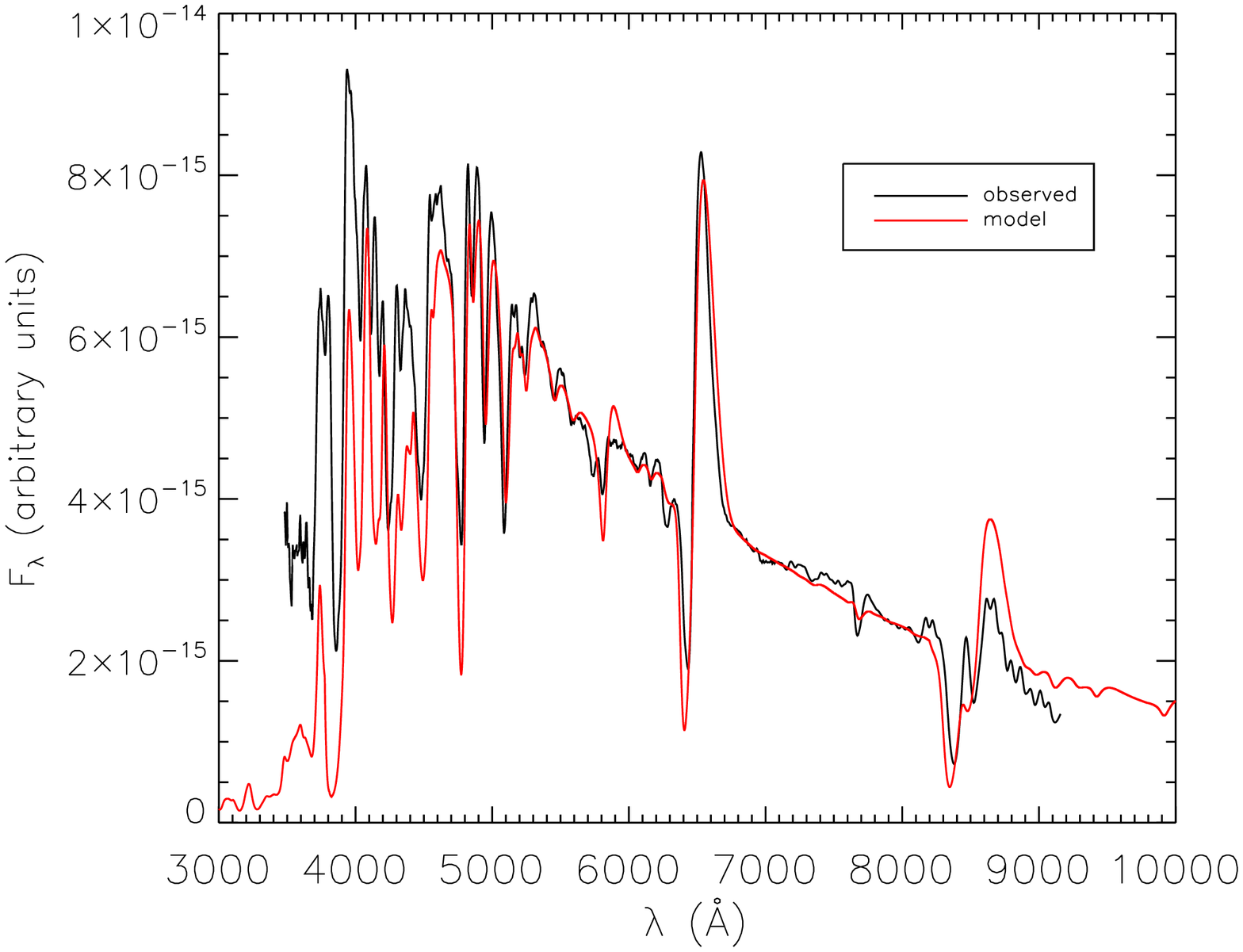}
\end{center}
\caption{\label{fig:na10_d17}. The observed spectrum on Jul 14,
  2005 is compared to a model spectrum using solar  abundances but
  sodium has been enhanced by a factor of 10. The
  observed spectrum has been 
  dereddened using $E(B-V) = 0.035$.}
\end{figure}

To examine the evolution of SN~2005cs we studied the spectrum obtained
on July 14, 2005. Using our results obtained above we will take the
fiducial reddening to be $E(B-V) = 0.035$. Figure~\ref{fig:d17_std}
(top panel)
shows our best fit model with solar abundances. The parameters for
this model were: \teff\ = 6000~K, \vno = 4000~\kmps, and $n = 8$. The
overall fit is quite good. The only noticeable mismatch is the trough
just blueward of 5700~\ang\ is not well fit. The red part of this
trough is almost certainly due to Na~I~D. \citet{pasto05cs06}
identify the feature just to the blue as due to Ba~II $\lambda$5854,
which won't be present in our models without enhanced s-process
abundances or considering NLTE effects of Ba~II, which is beyond the
scope of this work (see below).
The bottom panel of Figure~\ref{fig:d17_std} shows the fit with the same model
parameters but CNO enhanced abundances. There is no noticeable
degradation in the fit, and the enhanced sodium from CNO processing
leads to a feature in 
the right place to fit the red part of the feature near 6000~\ang. To
further test the sensitivity to the sodium abundance, we calculated a
model with solar abundances, but with sodium enhanced by a factor of
ten by number. Keeping the other parameters fixed, this model is shown
in Figure~\ref{fig:na10_d17}. Clearly the enhanced sodium leads to a
feature at the right place that is in fact too strong. Thus, there is
evidence for enhanced sodium which would most likely come from CNO
processing. Overall the sodium enhanced model is a bit too cool, but
we have held the other parameters fixed in order to illustrate just
the effect of enhancing the sodium abundance.

\clearpage

\section{Day 34}

\begin{figure}
\leavevmode
\begin{center}
\includegraphics[width=0.75\textwidth,angle=90]{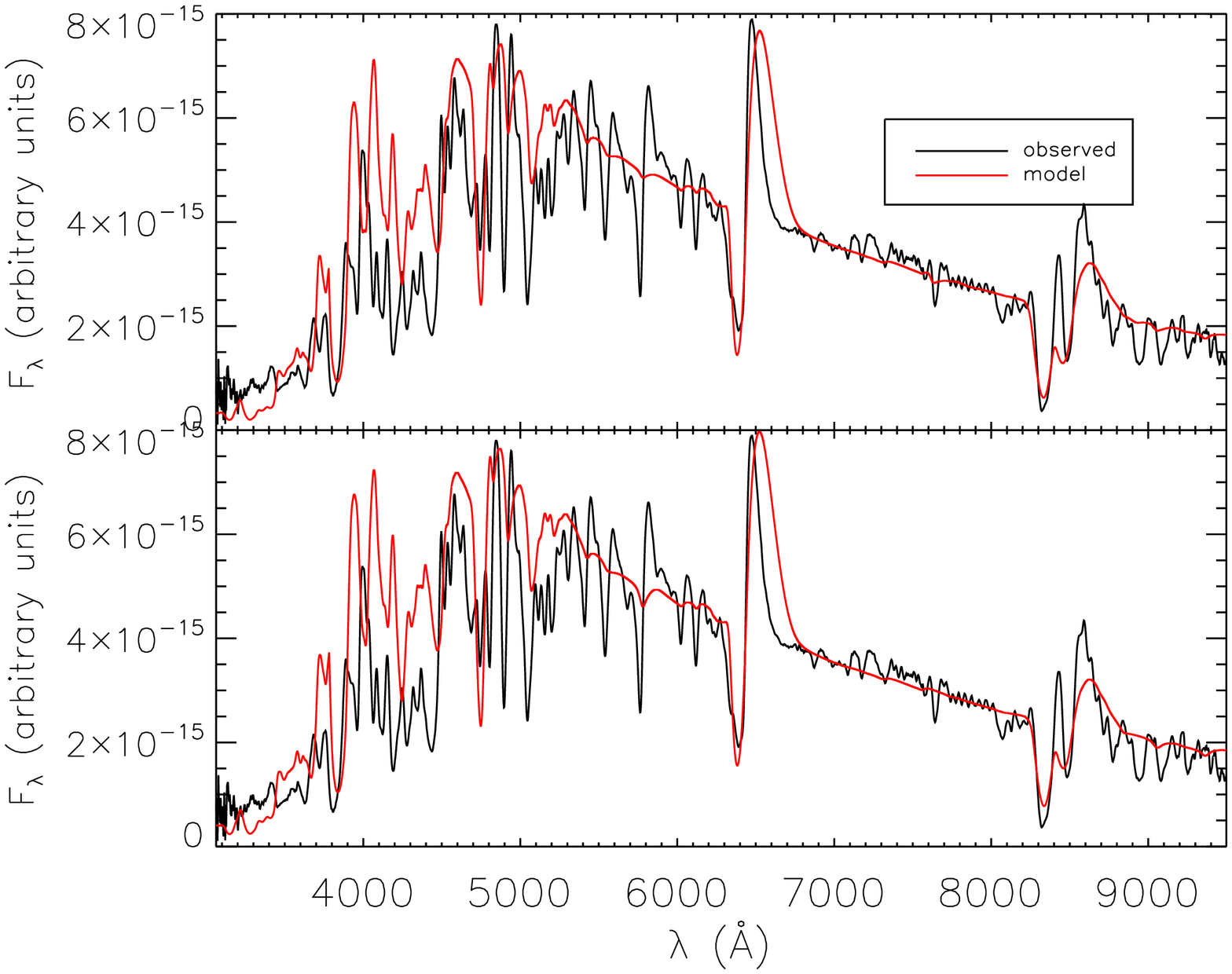}
\end{center}
\caption{\label{fig:d34_std}. The observed spectrum on Jul 31,
  2005 is compared to a model spectrum using solar abundances (top
  panel) and CNO enhanced abundances (bottom panel). The
  observed spectrum has been 
  dereddened using $E(B-V) = 0.035$.}
\end{figure}

\begin{figure}
\leavevmode
\begin{center}
\includegraphics[width=0.75\textwidth,angle=90]{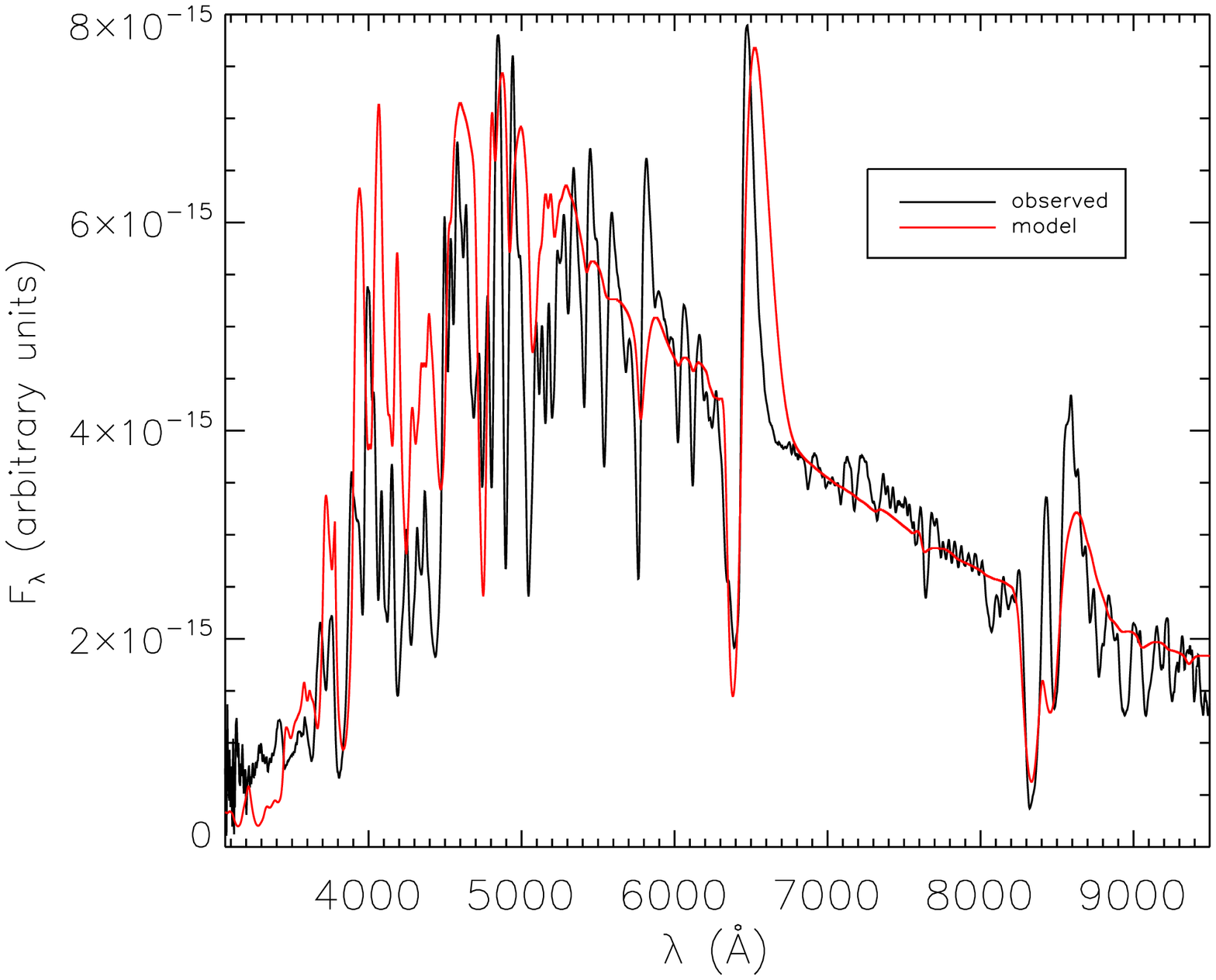}
\end{center}
\caption{\label{fig:na10_d34}. The observed spectrum on Jul 31,
  2005 is compared to a model spectrum using solar  abundances but
  sodium has been enhanced by a factor of 10. The
  observed spectrum has been 
  dereddened using $E(B-V) = 0.035$.}
\end{figure}

As our final epoch, we model spectra obtained on July 31, 2005. 
The parameters for
this model were: \teff\ = 5500~K, \vno\ = 6000~\kmps, and $n = 12$. This
parameter set is somewhat odd, considering that it implies that the
photosphere has actually moved outward from 4000~\kmps\ on day 17 to
6000~\kmps\ on day 31 and that the density profile is different. This
contradiction indicates the limits of our modeling with a simple
single power-law and uniform compositions. Figure~\ref{fig:d34_std}
(top panel)
shows the fit with solar abundances. Overall the color is well
reproduced, \halpha, and both the Ca II H+K and IR triplet features are
well fit, but many other lines are not, including other lines in the
Balmer series. In fact while the \halpha\ absorption trough is quite
well fit, the emission peak is too wide and too red. This is almost
certainly due to our simple single power-law density profile as we
will discuss in future work. Figure~\ref{fig:d34_std}
(bottom panel) shows the fit with CNO processed abundances. As was the
case for day 
17, the only real improvement is that now Na~I~D appears in the
synthetic spectrum, although it is too weak. Figure~\ref{fig:na10_d34}
shows the result of enhancing Na by a factor of ten in the solar
abundance mixture, now Na~I~D shows up nicely, although not quite at
the right velocity or strength. 

\begin{figure}
\leavevmode
\begin{center}
\includegraphics[width=0.75\textwidth,angle=90]{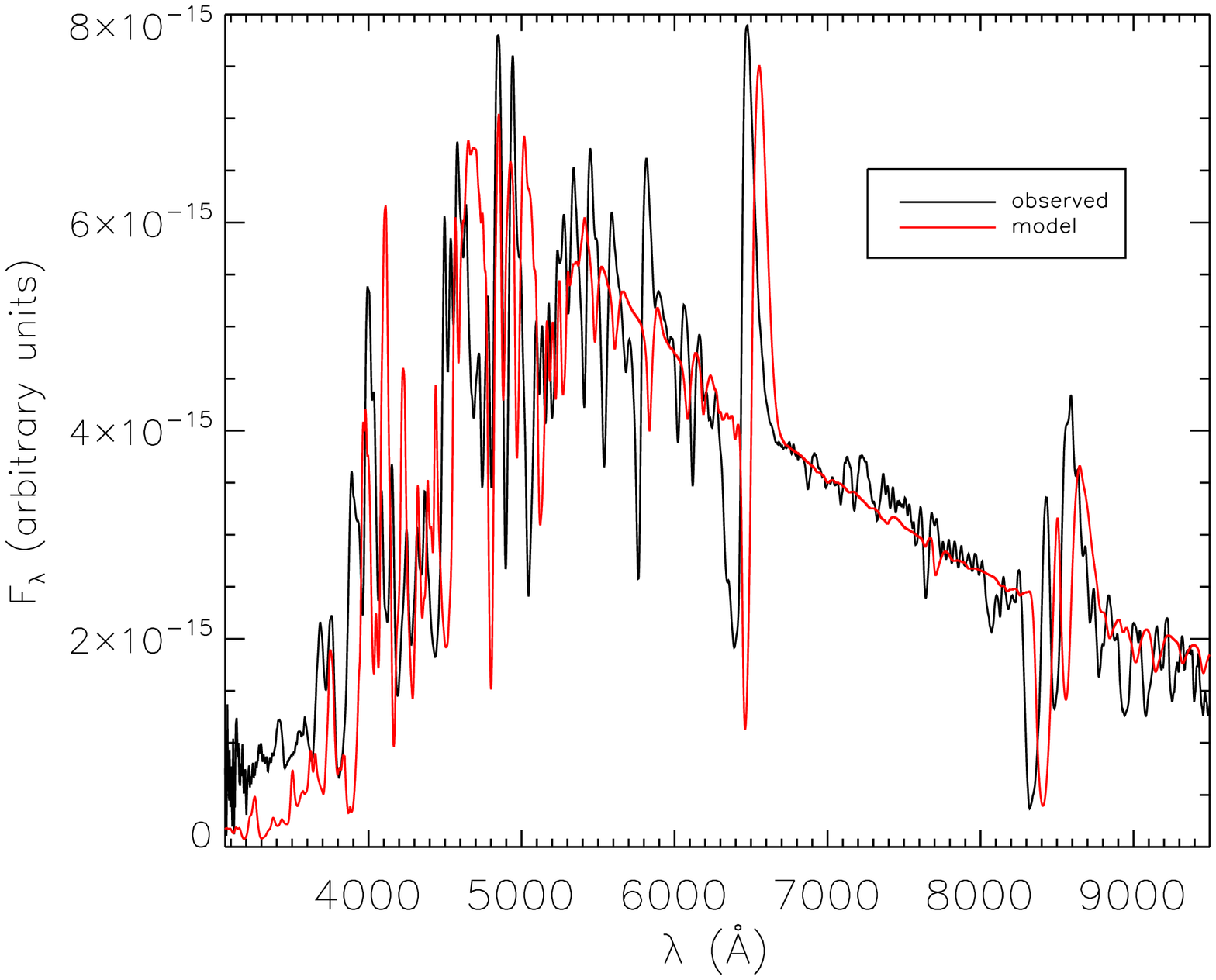}
\end{center}
\caption{\label{fig:sol_v3000_d33}. The observed spectrum on Jul 31,
  2005 is compared to a model spectrum using solar abundances and
  $\vno 
  = 3000$~\kmps. The
  observed spectrum has been 
  dereddened using $E(B-V) = 0.035$.}
\end{figure}

Figure~\ref{fig:sol_v3000_d33} shows a model with solar abundances,
\teff\ = 5500~K, \vno\ = 3000~\kmps, and $n = 12$. While narrow lines
appear, neither the Balmer lines nor the calcium lines are well fit and
none of the absorption troughs appear at the right velocity. 

\citet{pasto05cs06} found evidence for
s-process elements such as barium, and charged particle elements such
as scandium, which if present in
our models would certainly improve the fit in
Fig.~\ref{fig:d34_std}. In our models these species are treated in
LTE, but it is possible that enhanced abundances would be required to
produce the strong lines observed. In studying the SN~1987A analog
SN~1998A \citet{pasto98A05} concluded that the barium lines are quite
temperature sensitive and thus the difference in strength between
these lines and the very strong lines observed in 1998A could well be
due to NLTE effects. \citet{utchug05} also found that the barium lines
in SN~1987A was strongly dependent on the radiation field and that a
full NLTE solution with a reduced UV flux due to line-blanketing
reduces the need for enhanced barium abundance. Thus, with a better
model for this epoch and treating barium and scandium in NLTE we are
likely to get significantly improved fits, without required enhanced
s-process abundances. This is a subject for future work.

\clearpage

\clearpage
\section{Distance}

One method of determining distances using Type II
supernovae is the ``expanding photosphere method''
\citep[EPM,][]{kkepm,branepm,eastkir89,skeetal94,esk96,hamuyepm01,leonard99em02} a variation of the
Baade-Wesselink method \citep{baadeepm}. The EPM method assumes
that for SNe~IIP, with intact hydrogen envelopes, the spectrum is
not far from that of a blackbody and hence the luminosity is
approximately given by
\[
L = 4\pi\,\zeta^2\,R^2\,\sigma\,T^4
\]
where $R$ is the radius of the photosphere, $T$ is the effective
temperature,  $\sigma$ is the radiation constant, and $\zeta$ is the
``dilution factor'' which takes into account that in a scattering
dominated atmosphere the blackbody is
diluted \citep*{hlw86a,hlw86b,hw87}. The temperature is found
from observed colors, so in fact is a color temperature and not an
effective temperature, the photospheric velocity can be estimated from
observed spectra using the velocities of the weakest lines, 
\[ R = v\, t, \]
the dilution factor is estimated from synthetic spectral models,
and $t$ comes from the light curve and demanding self-consistency.

Both an advantage and disadvantage of EPM is that it primarily requires
photometry. Spectra are only used to determine the photospheric
velocity, colors yield the color temperature, which in turn is used to
determine 
the appropriate dilution factor (from model results).
This method suffers from uncertainties in determining the dilution
factors, the difficulty of knowing which lines to use as velocity
indicators, uncertainties between color temperatures and effective
temperatures, and questions of how to match the photospheric radius
used in the models to determine the dilution factor and the radius of
the line forming region \citep{hamuyepm01,leonard99em02}.  In spite of
this the EPM method was successfully applied to SN~1987A in the
LMC \citep{eastkir89,bran87a} which led to hopes that the EPM method
would lead to accurate distances, independent of other astronomical
calibrators. Recently, the EPM method was applied to the very well
observed SN~IIP
1999em \citep{hamuyepm01,leonard99em02,elmhhamdietal99em03}. All three
groups found a distance of 7.5--8.0~Mpc. 
\citet{leonard99em03} subsequently used \emph{HST} to obtain a
Cepheid distance to the parent galaxy of SN~1999em, NGC 1637, and found
$11.7 \pm 1.0$~Mpc, a value 50\% larger than that obtained with
EPM. \citet{dessart06a} calculated updated values for the dilution
factor and found good agreement with the Cepheid distance using EPM,
but emphasized that dilution factors would have to be calculated for
each supernova individually. Using SEAM \citet{bsn99em04} found a
distance to SN~1999em of $12.5 \pm 2.3$~Mpc.

With modern detailed NLTE radiative transfer codes, accurate synthetic
spectra of all types of supernovae can be calculated.  The
\textbf{S}pectral-fitting \textbf{E}xpanding \textbf{A}tmosphere
\textbf{M}ethod  \citep[SEAM,][]{b93j3,b94i1,l94d01,mitchetal87a02} was
developed using the generalized stellar atmosphere code
\texttt{PHOENIX} 
\citep[for a review of the code see][]{hbjcam99}. While SEAM is 
similar to EPM in spirit, it avoids the use of dilution factors
and color temperatures. Velocities are determined accurately by
actually fitting synthetic  and observed spectra. The radius is
still determined by the relationship $R = vt$, (which is an excellent
approximation because all supernovae quickly reach homologous
expansion) and the explosion time is found by demanding self consistency.
SEAM uses all the spectral information available in the observed spectra
simultaneously which broadens the base of parameter determination.
Since the spectral energy distribution is known completely from the
calculated synthetic spectra, one may calculate the absolute
magnitude, $M_X$,
in any photometric band $X$, 
\[
M_{X} = -2.5 \log \int_{0}^{\infty} S_{X}(\lambda)\, L_{\lambda}\,
d\lambda + C_X \,
\]
where $S_{X}$ is the response of filter $X$, $L_{\lambda}$ is the
luminosity per unit wavelength, and $C_X$ is the zero point of filter
$X$ determined from standard stars. Then one immediately obtains a
distance modulus $\mu_X$, which is a measure of the distance
\[ \mu_X \equiv m_X - M_X - A_X - K_x = 5\log{(d/10\textrm{pc})}, \]
where $m_X$ is the apparent magnitude in band $X$, $A_X$ is the
extinction due to dust along the line of sight both in the host
galaxy and in our own galaxy, and $K_X$ is the k-correction
\citep{okesand_kcorr68}.  The SEAM
method does not need to invoke a blackbody assumption or to calculate
dilution factors.

We calculate a SEAM distance to SN~2005cs using our first two early
dates, day 5 and day 17, since our best fit to day 34 was
contradictory in that the photosphere moved out and the density slope
was different at a previous time. With only two days we don't have
much leverage on the time of explosion, thus we adopt the explosion
time of \citet{pasto05cs06}. Using 5 photometric bands $UBVRI$ and our
adopted reddening of $E(B-V) = 0.035$ we find $\mu = 29.5 \pm 0.18$
which corresponds to a distance of $7.9^{+0.7}_{-0.6}$~Mpc where the error
is just the formal $1-\sigma$ error and does not include systematic
error. Using the EPM method \citet{tv05cs06} found $D=7.59 \pm 1.02$~Mpc,
and using a standard candle method \citep{hamuy03a,nugent06} they
found $D=6.02 \pm 1.3$~Mpc. Using the SEAM method on SN~1994I a Type
Ic that is also in M51 we found $D=6.02 \pm 1.92$~Mpc
\citep{b94i1}.  Given that SNe~Ic are much more difficult to model
than SNe~II, the fact that the two SEAM results agree within the
errors is strong evidence that not only is the SEAM method robust, but
the errors can be reasonably well estimated.  \citet{tv05cs06}
summarize the other distance estimates obtained to M51 and find an
average value of $D=7.1 \pm 1.2$~Mpc. Given the wide range of
distances obtained and the relative closeness of M51, an effort should
be made to obtain a Cepheid distance to M51.

\section{Discussion and Conclusions}

We have calculated a small grid of simple parameterized explosion
models for SN~2005cs. We have shown that the reddening estimated in
previous studies is likely too high and we adopted the lowest possible
value, that of foreground reddening alone. Additionally we have shown
that while there is evidence for enhanced nitrogen and sodium
as would be expected from CNO processing in the pre-supernova
envelope, the specific set of abundances that we adopted were too
depleted in oxygen to match the observed spectrum. We have not
attempted to exhaust the full parameter space and determine exact
abundances in this work, merely to show that such an approach is
feasible given sufficient computational resources. Additionally, we
had trouble obtaining a physically plausible fit to the final epoch we
studied, 34 days past explosion. All of these shortcomings point that
a better approach would be to run large grids using the best current
stellar evolution models and compare the results to many well-observed
SNe~IIP. This should be feasible in the near future. Using the SEAM
method we obtained a distance to M51 which is somewhat larger than the
accepted distance. 
The SEAM method has been shown to be reliable in modeling SNe~IIP and
our result is reliable enough that it should be calibrated with a
Cepheid distance to M51. This could then be compared with the results
from a larger more homogeneous grid.

\acknowledgements
  We think Andrea Pastorello for providing us with unpublished spectra
  and Abouazza Elmhamdi and Andrea Pastorello for helpful discussions.
  This work was supported in part by by NASA grants NAG5-3505 and
  NAG5-12127, and NSF grant AST-0307323.  PHH was supported in part by
  the P\^ole Scientifique de Mod\'elisation Num\'erique at ENS-Lyon.
  This research used resources of the National Energy Research
  Scientific Computing Center (NERSC), which is supported by the
  Office of Science of the U.S.  Department of Energy under Contract
  No. DE-AC03-76SF00098; and the H\"ochstleistungs Rechenzentrum Nord
  (HLRN).  We thank all these institutions for a generous allocation
  of computer time.


\end{document}